\documentclass[%
 reprint, superscriptaddress,
 amsmath,amssymb,
]{revtex4-2}

\usepackage[english]{babel}
\usepackage{graphicx}
\usepackage{siunitx}
\usepackage{hyperref}
\usepackage{bm}

\DeclareSIUnit\electronmass{\text{\ensuremath{m}}_{\mathrm{e}}}
\DeclareSIUnit\angstrom{\text{\AA}} 
\DeclareSIUnit{\rpm}{rpm}
\DeclareSIUnit\bar{bar}
\DeclareMathAlphabet\mathbfcal{OMS}{cmsy}{b}{n}
\newcommand{\appref}[1]{Appendix~\ref{#1}}

\begin{document}

\preprint{APS/123-QED}

\title{\texorpdfstring{De Haas - van Alphen study of the Dirac\\ nodal-line semimetal candidate TaPtTe$_5$}{}}

\author{Maximilian Daschner}
\email{md867@cam.ac.uk}
\email{maximilian.daschner@lmu.de}
\affiliation{Cavendish Laboratory, University of Cambridge, Cambridge CB3 OHE, United Kingdom}
\affiliation{Fakultät für Physik, Ludwig-Maximilians-Universität, München, Germany}

\author{F. Malte Grosche}
\email{fmg12@cam.ac.uk}
\affiliation{Cavendish Laboratory, University of Cambridge, Cambridge CB3 OHE, United Kingdom}


\begin{abstract}
We report a quantum oscillation study in the Dirac nodal-line semimetal candidate TaPtTe$_5$. The Fermi surface is probed via magnetic torque measurements with the magnetic field applied in the crystallographic \textit{a-b} and \textit{b-c} planes. The experimentally determined de Haas - van Alphen frequencies are consistent with results from band structure calculations. This study serves as an extension to the scarce quantum oscillation data on TaPtTe$_5$ currently present in the literature.
\end{abstract}

\maketitle

\section{Introduction}
Ever since the discovery of Dirac and Weyl materials, interest in more exotic topological semimetals has surged. This led to thorough theoretical and experimental studies of various different types of non-trivial band crossings. Among them, the simplest nodal-point Hamiltonian can be expressed through a spin-$\frac{1}{2}$ representation with linear dispersion \cite{soluyanov2015type}. Higher-dimensional representations of SU(2) are in principle also possible, and the polynomial order can be quadratic or cubic around the touching point \cite{robredo2024multifold,tang2017multiple,lv2021experimental}. Candidates for Dirac and Weyl semimetals are e.g. layered transition metal tellurides such as the family XYTe$_4$ (X = Nb, Ta; Y = Ir, Rh) \cite{koepernik2016tairte,liu2017van,haubold2017experimental,zhou2018coexistence,khim2016magnetotransport,schonemann2019bulk,zhou2019nonsaturating}.

More recently, so-called nodal-line semimetals (NLSM) have gained attention as well. Such NLSMs form band crossings along a line instead of just certain points in reciprocal space. The structural similarities and intricate topological properties present in ternary transition metal tellurides motivate investigation of such systems further. A fairly novel candidate that falls into that category is the in-plane anisotropic non-magnetic family of compounds TaXTe$_5$ (X = Ni, Pd, Pt) \cite{daschner2025mechanical,daschner2024probing,xu2020anisotropic,wang2022coexistence,liimatta1989synthesis,hu2022transport,ma2023quasi,ye2022anisotropic,huang2023magnetic,li2022coexistence,chen2021three,hao2021multiple,zhou2024origin,lu2022topologically,jiao2020topological,mar1991synthesis,xiao2022dirac,jiao2021anisotropic}. Shubnikov - de Haas (SdH) and de Haas–van Alphen (dHvA) oscillations reveal non-trivial Berry phases and low effective masses, which provide compelling experimental evidence for the existence of Dirac fermions in this family of compounds.

In this work, we examine one representative of this family, TaPtTe$_5$. DHvA oscillations in TaPtTe$_5$ have been previously reported for fields along the crystallographic \textit{b}-axis \cite{jiao2021anisotropic}, but a more comprehensive determination of its electronic structure has been lacking. Here, we report a full rotation study of dHvA oscillations in TaPtTe$_5$ for fields in the \textit{a}-\textit{b} and \textit{b}-\textit{c} planes. Our electronic structure calculations within density functional theory (DFT) suggest four-fold degenerate nodal-lines in the $k_z = \pm\pi/c$ plane of the Brillouin zone. We compare the dHvA oscillation frequencies extracted from the numerically predicted electronic structure to experimentally determined dHvA frequencies. Our findings suggest that the Fermi surface of TaPtTe$_5$ includes a small cylindrical pocket that encloses a nodal line in the $k_z = \pm\pi/c$ plane.

\begin{figure*}[t!]
  \centering
  \includegraphics[width=\textwidth]{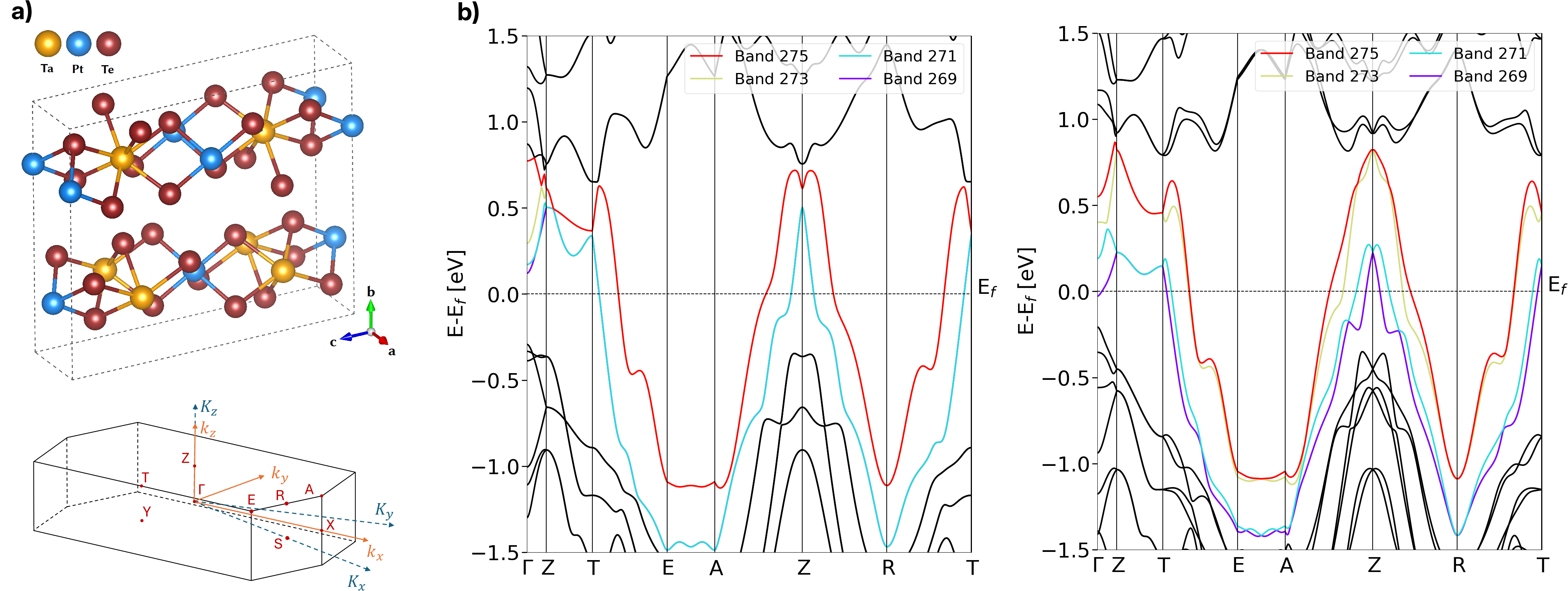}
  \caption{a) Crystal structure of TaPtTe$_5$ where the unit cell is indicated by dashed lines, and the corresponding Brillouin zone. Lower case $k_n$ ($n= x, y, z$) point in conventional unit cell vector directions, while upper case $K_n$ are the reciprocal lattice vectors that represent the periodicity of the Brillouin zone. b) Band structure without (left) and with (right) spin-orbit coupling in TaPtTe$_5$ with all high-symmetry points chosen in the plane at which $k_z = \pm\pi/c$ (except $\Gamma$). Spin-degenerate partners for each band are labelled with even numbers (not shown here). When SOC is excluded, all bands show four-fold degeneracy, and the $k_z = \pm\pi/c$ plane can thus be viewed as a nodal surface. Upon including SOC, the degeneracy is lifted (spin degeneracy remains) in most of the $k_z = \pm\pi/c$ plane, but remains along the Z-T high-symmetry line and in the high-symmetry point R. The observations are independent of the energy, i.e. nodal-lines and nodal-points form at all energies in these high-symmetry regions.}
  \label{TaPtTe5_crystalstructure}
\end{figure*}

\section{Methods}

Single crystals of TaPtTe$_5$ were grown via the self-flux growth method using Te as flux following S. Xiao et al. \cite{xiao2022dirac}. Ta powder (Alfa Aesar: 99.95\%), Pt fragments cut off from an ingot (Produits Artistiques Métaux Précieux: 99.95\%) and Te lumps (Alfa Aesar: 99.999\%) were mixed together at a molar ratio of Ta:Pt:Te = 1:1:10 inside an Ar-filled glove box. The mixture is placed inside an alumina crucible which is then put inside an evacuated quartz ampoule. It is then heated in a muffle furnace with the temperature profile described in \cite{jiao2021anisotropic,xiao2022dirac}, after which crystals are extracted mechanically. The energy-dispersive X-ray spectrum (EDS) and a scanning electron microscope (SEM) image of a typical crystal are shown in \appref{Characterisation}.

Among the crystals in each growth, the majority are alien phases such as TaTe$_2$, TaTe$_4$, TaPt$_2$, PtTe$_2$ or Te as reported in the literature \cite{mar1991synthesis}, where only a small fraction ($\sim1\%$) of the crystals in the batch turned out to be TaPtTe$_5$. EDS was used to distinguish the latter from other phases, however the shape was often enough to visually filter out most unwanted crystals. The obtained TaPtTe$_5$ crystals are air-stable and grow in shiny grey flattened needle-like shapes. The residual resistance ratio (RRR) resulted in $\rho(\SI{298}{\kelvin})/\rho(\SI{1.6}{\kelvin})\approx 13$ which is close to the value of RRR $\approx 15$ reported before \cite{jiao2021anisotropic}.

Magnetic torque measurements were conducted with custom-made piezoresistive cantilevers read out through a Wheatstone bridge via a lock-in amplifier. Cryogenic temperatures were reached using a custom-designed cryostat by Oxford Instruments with fields of up to \SI{15}{\tesla}. All measurements were performed at the base temperature of \SI{1.6(1)}{\kelvin}.

To gain insights into the electronic structure of TaPtTe$_5$, we performed DFT calculations using Wien2k \cite{blaha2001wien2k} with the same parameters as chosen in \cite{daschner2024probing}. The generalised gradient approximation (GGA) based on the PBE exchange-correlation potential \cite{perdew1996generalized} was used. The atomic sphere radii (muffin-tin radii) were chosen as $R_{mt} = 2.5$ a.u. for Ta and Pt atoms and $R_{mt} = 2.43$ for all Te atoms. The plane-wave cut-off parameter was chosen as  $R_{mt}K_{max} = 8$, where $R_{mt}$ is the smallest atomic sphere radius in the unit cell and $K_{max}$ is the magnitude of the largest $k$-vector. Calculations were performed on a $59 \times 59 \times 13$ k-point mesh in the full Brillouin zone for the band structure calculations shown here. DHvA frequencies were extracted using the Supercell K-space Extremal Area Finder (SKEAF) \cite{rourke2009electronic,julian2012numerical}, while Fermi surface plots were generated with \textit{py\_FS} \cite{Weinberger2024fsplot}.

\section{\textit{Ab initio} calculations}

TaPtTe$_5$ crystallises in an orthorhombic layered structure with space-group Cmcm (No. 63), with lattice constants \textit{a} = \SI{3.729(4)}{\angstrom}, \textit{b} = \SI{13.231(11)}{\angstrom}, and \textit{c} = \SI{15.452(10)}{\angstrom} as shown in \autoref{TaPtTe5_crystalstructure} a), and was reported to exhibit Pauli paramagnetism \cite{mar1991synthesis}. The unit cell consists of two TaPtTe$_5$ layers, where each layer can be described as a series of bicapped trigonal prismatic TaTe$_5$ chains stuck together with Pt atoms. The TaPtTe$_5$ layers are stacked by shifting them along the \textit{a}-axis by half a lattice vector with respect to each other.

The numerical analysis here follows \cite{xiao2022dirac}, and will be used to interpret experimental results from quantum oscillation measurements. \autoref{TaPtTe5_crystalstructure} b) illustrates the resulting band structure without and with spin-orbit coupling (SOC) with all high-symmetry points chosen in the plane at which $k_z = \pm\pi/c$ (except $\Gamma$). In the case without spin-orbit coupling, one can clearly see two-band degeneracies (fourfold if spin for each band is included) whose degeneracy is only lifted away from the $k_z = \pm\pi/c$ plane as illustrated on the $\Gamma$-Z high-symmetry line. The $k_z = \pm\pi/c$ plane can therefore be referred to as a nodal surface. Here the labels only include one band since their spin-degenerate partners are identical due to the presence of $\mathcal{PT}$ symmetry across the whole Brillouin zone. When SOC is included, the two-band degeneracies in this plane are mostly lifted (spin degeneracy still remains) except at the high-symmetry line Z-T and in the high-symmetry point R. To get a better overview of all the nodal lines away from the high-symmetry regions, we visualise the four bands that cross the Fermi energy and hence contribute Fermi pockets to dHvA oscillations in \appref{nodal_line_plane} for the whole $k_z = \pm\pi/c$ plane (including spin-orbit coupling). The energetic difference between those bands clearly shows the presence of band crossings. The nodal line along Z-T is protected from spin-orbit coupling by the crystal symmetries, which was shown theoretically for this and other materials in the same symmetry group (Cmcm), see e.g. TaNiTe$_5$ \cite{hao2021multiple}, TaPtTe$_5$ \cite{xiao2022dirac}, LaNiGa$_2$ \cite{badger2022dirac}, and Th$_2$BC$_2$ \cite{wang2022symmetry}.

\section{Results and Discussion}
Quantum oscillation measurements in the magnetic torque $\tau$ in an external field $\mathbf{B}$ can be captured via the relation \cite{shoenberg2009magnetic}

\begin{equation}\label{torque_LK}
	\resizebox{0.9\linewidth}{!}{$\tau = C B^{3/2}\sum\limits_{\text{i}} \frac{dF_i}{d\theta} \left|\frac{\partial^2 A_i}{\partial k^2_{\parallel}}\right|^{-1/2} \sum\limits_{p=1}^{\infty}p^{-3/2}R_D R_T R_S \sin\left(2\pi p \left(\frac{F_i}{B}-\gamma\right)\pm\delta\right)$}
\end{equation}
where $C$ is a field-independent constant, $\theta$ is the angle between the extremal area $A$ and the magnetic field, while $k_{\parallel}$ is the wavevector parallel to $\mathbf{B}$. The indices $i$ and $p$ range over all orbits and higher harmonics, respectively. $R_D$, $R_T$, and $R_S$ are various damping terms arising from finite electron lifetime, finite temperature, and spin-splitting, respectively. The phase shift $\gamma$ arises from the topological properties of the semimetal and is often approximated with $\gamma = \frac{1}{2}-\frac{\phi_B}{2\pi}$ with Berry phase $\phi_B$, while $\delta$ depends on the dimensionality of the Fermi sheet. To find the frequency $F$ one can use the Onsager relation
\begin{equation}\label{Onsager_relation}
	F = \left(\frac{\hbar}{2\pi e}\right) A
\end{equation}
where $A$ is the extremal area of an orbit, while $\hbar$ and $e$ are the reduced Planck constant and the electron charge, respectively.

\subsection{DHvA frequencies}
DHvA oscillations in the magnetisation were investigated by W.-H. Jiao et al. \cite{jiao2021anisotropic} up to \SI{7}{\tesla} with the magnetic field along the crystallographic \textit{b}-axis. The authors performed a mass study and found two fundamental frequencies at F = \SI{63.5}{\tesla} and F = \SI{95.2}{\tesla} with light effective masses of m = \SI{0.108}{\electronmass} and m = \SI{0.119}{\electronmass}, respectively. Although DFT calculations were also performed, the dHvA data consisting of only two frequencies along one crystallographic axis is not sufficient to verify such calculations with high accuracy, and thus a more thorough investigation to higher fields, and along multiple crystallographic axes is required.

\begin{figure}[t]
  \centering
  \includegraphics[width=\columnwidth]{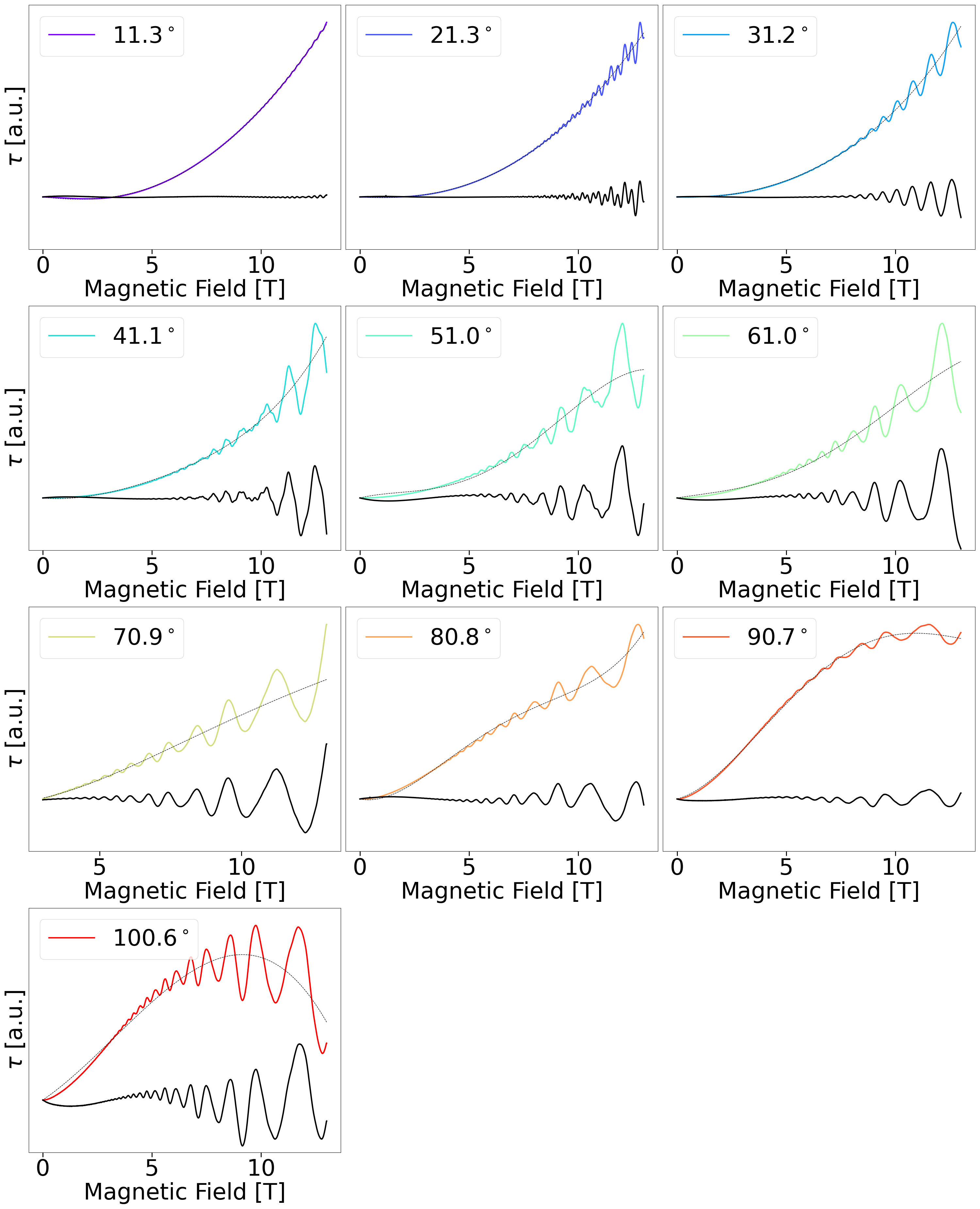}
  \caption{Magnetic torque of a TaPtTe$_5$ sample rotated in the \textit{b}-\textit{c} plane measured up to \SI{13}{\tesla}. The whole data set was fitted with a fourth-degree polynomial (dashed black line), and the background-subtracted data is given as a solid black line. \SI{0}{\degree} corresponds to the magnetic field being aligned with the \textit{c}-axis.}
  \label{TaPtTe5_overall_torque_individual_b-c_2}
\end{figure}

An overview of magnetic torque measurements performed here with the external magnetic field $\mathbf{B}$ applied in the \textit{b}-\textit{c} plane is shown in \autoref{TaPtTe5_overall_torque_individual_b-c_2}. The non-linear background, mostly induced by the inhomogeneity of the piezoresistive cantilever, is corrected by subtracting a fourth-degree polynomial. In \appref{raw_data_dhva}, additional data obtained on a different cantilever with the magnetic field of up to \SI{15}{\tesla} is provided together with respective fast Fourier transforms (FFT). The raw data for field sweeps with $\mathbf{B}$ in the \textit{a}-\textit{b} plane and its respective FFT spectrum are also included.

Due to the lack of data at temperatures above the base temperatures of the cryostat, no effective masses or Dingle temperatures could be extracted reliably. The angles were determined from the torque caused by the weight of the sample in zero field. The extracted frequency peaks are shown in \autoref{TaPtTe5_frequencies_torque}. Note that quantum oscillations in the \textit{a-b} and \textit{b-c} planes were measured separately. In each plane, frequencies measured at angles below \SI{0}{\degree} and above \SI{90}{\degree} were folded back into the \SI{0}{\degree}-\SI{90}{\degree} range due to the periodicity of the crystal.

\begin{figure}[t]
  \centering
  \includegraphics[width=\columnwidth]{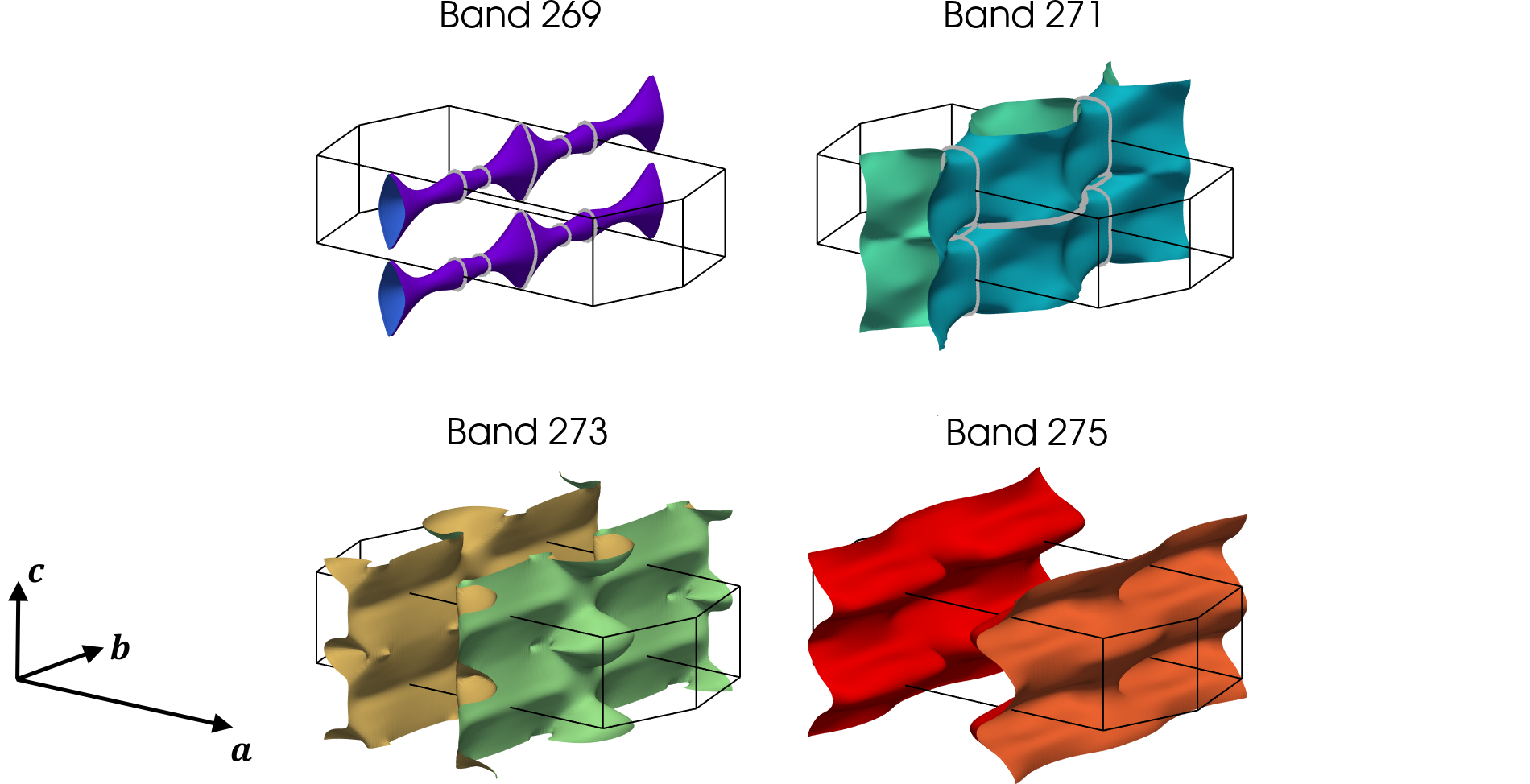}
  \caption{Fermi sheets for TaPtTe$_5$ with the Fermi level shifted to lower energies by \SI{20.4}{\milli\electronvolt}. Extremal orbits are shown for the magnetic field along $a$, $b$ and $c$ as grey lines in the first Brillouin zone.}
  \label{TaPtTe5_shifted_FS}
\end{figure}

To compare these experimental findings with results obtained from DFT calculations, the band structure can be used to extract the Fermi surface as shown in \autoref{TaPtTe5_shifted_FS}. Extremal areas with respect to the magnetic field direction are indicated by grey lines and allow one to find the frequency expected from dHvA oscillations via the Onsager relation. Only two bands (band 269 and 271) possess closed orbits and contribute dHvA frequencies. A good match between calculated and observed dHvA frequencies was obtained by moving the band edges of band 269 and 271 by \SI{20.4}{\milli\electronvolt}. The other two bands (273 and 275) may shift in the opposite direction for overall charge neutrality, or slight off-stoichiometry could cause a change in band filling. The extracted frequencies as a function of angle are included in \autoref{TaPtTe5_frequencies_torque} and show consistency with the experimental values. Band 269 contributes multiple hole-like frequencies due to its irregular shape, which diverge as the magnetic field deviates from the \textit{b}-axis in line with the behaviour expected from quasi-cylindrical Fermi surfaces. This Fermi sheet is particularly interesting because it wraps around the nodal line Z-T in the $k_z=\pm\pi/c$ plane. Band 271 shows an electron-like frequency branch with its centre on the \textit{a}-axis, and a hole-like frequency branch with its centre on the \textit{b}-axis.

Overall, the dHvA oscillations seem to agree well with the DFT calculations, which confirms the numerical results that form the basis for \appref{nodal_line_plane} and provides indirect evidence for the existence of such Dirac nodal-lines.

\subsection{Phase analysis}
The phase shift measured in quantum oscillations can in principle allow one to obtain information about the topological properties of a material \cite{mikitik1999manifestation}, if it can be extracted reliably and other non-topological contributions can be ruled out.

\begin{figure}[t]
        \includegraphics[width=\linewidth]{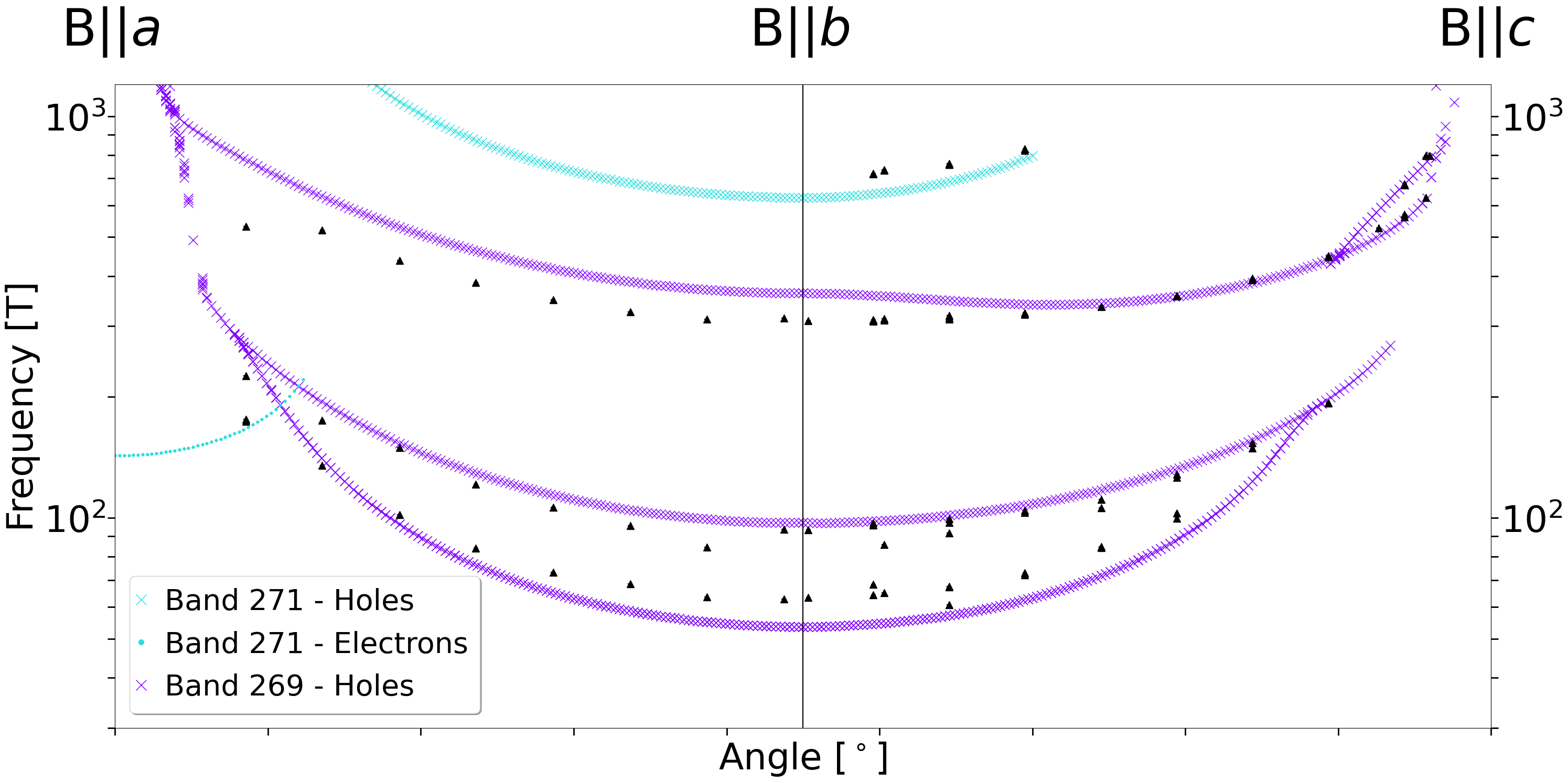}
	\caption{De Haas - van Alphen frequencies in the magnetic torque as a function of angle in the \textit{a-b} and \textit{b-c} planes for a TaPtTe$_5$ sample (black triangles). Results from DFT calculations are included as coloured points (electron-like orbit) and crosses (hole-like orbit). Ticks on the x-axis correspond to steps of \SI{20}{\degree}.}\label{TaPtTe5_frequencies_torque}
\end{figure}

Regarding the phase shifts in TaPtTe$_5$, we expect to obtain a non-zero Berry phase for the Fermi surface arising from band 269 as it surrounds the nodal-line along Z-T \cite{li2018rules}. All three branches from band 269 shown in \autoref{TaPtTe5_frequencies_torque} should in principle return such a non-trivial phase shift when the magnetic field is aligned with the \textit{b}-axis. W.-H. Jiao et al. \cite{jiao2021anisotropic} did indeed find non-zero Berry phase shifts for the lowest two hole-branches of band 269. It is however questionable how accurate these results are, since the magnetic fields used to measure dHvA oscillations did not exceed \SI{7}{\tesla} and are thus far below the quantum limit even for the lowest frequency at \SI{63.5}{\tesla}. The quantum limit is reached when the external magnetic field approaches the frequency of a cyclotron orbit. The uncertainty regarding the phase shift is evident in the fact that direct fits of the quantum oscillation amplitude and evaluation of Berry phase shifts using Landau fan diagrams are inconsistent with each other in the study performed by these authors.

Furthermore, \cite{jiao2021anisotropic} find different phase shifts for the two frequency branches, despite the fact that both encircle the same nodal-line and arise from the same band. Besides that, Berry phases are difficult to interpret as other, non-topological, effects can also lead to a phase shift of $\pi$, see e.g. spin-splitting \cite{wang2018vanishing} and other effects \cite{alexandradinata2023fermiology}.

Attempts to extract the phase shift from the data presented in this work shows similar uncertainty. Besides the low magnetic fields used in the measurements, the multi-oscillation nature of TaPtTe$_5$ introduces additional difficulty when extracting such phases. To determine the topological nature of this material, higher magnetic fields are therefore necessary.

\section{Conclusion}
In this work we presented numerical results that show the existence of four-fold degenerate Dirac nodal-lines in the $k_z=\pm\pi/c$ plane in the Brillouin zone of TaPtTe$_5$. After including spin-orbit coupling, this degeneracy is mostly lifted except at the high-symmetry line Z-T among others. De Haas - van Alphen frequencies extracted from magnetic torque measurements allow to obtain a complete picture of the Fermi surface and hence indirectly confirm the band structure in this material. Particularly the shape and size of the Fermi sheet arising from band 269 can be confirmed. Although the Berry phases can not reliably be extracted from the data presented here or in the literature, high-field measurements should in principle be able to reach the quantum limit of the lowest frequency branch arising from band 269, and confirm the presence of non-trivial Berry phases, and other signatures often found in Dirac semimetals.

\section{Acknowledgements}
The authors thank Ivan Kokanović for many insightful discussions.

\clearpage

\appendix

\section{Characterisation}\label{Characterisation}

\begin{figure}[ht]
    \includegraphics[height=3.3cm]{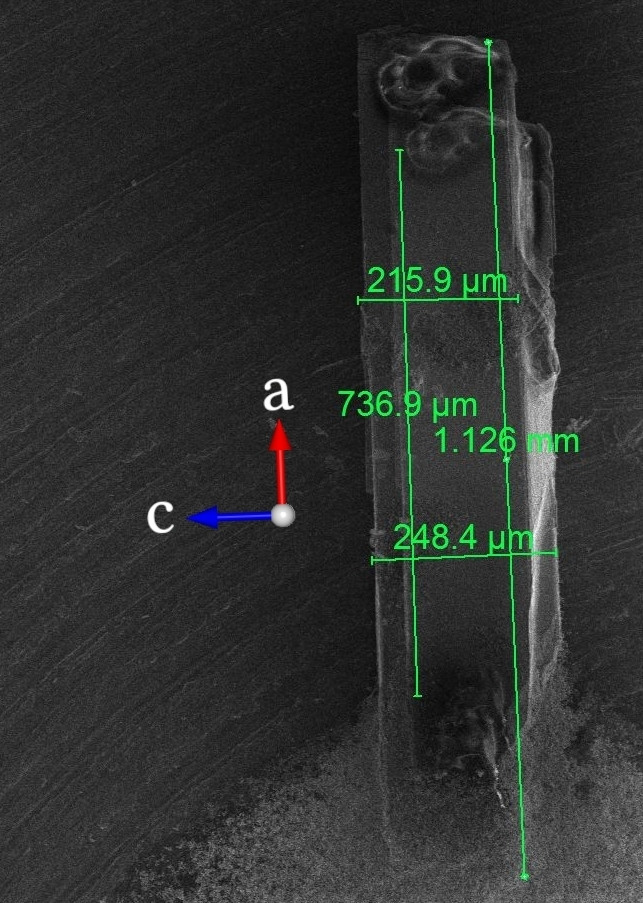}
    \includegraphics[height=3.3cm]{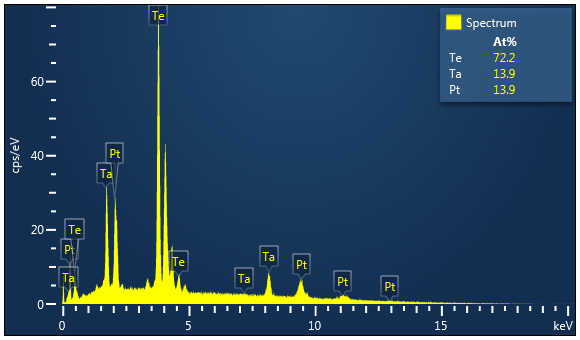}
    \caption{Left: Close-up scan in a SEM of a TaPtTe$_5$ sample. The sample is attached to a metallic pin with conductive silver epoxy at the bottom such that no charging effects occur during SEM operation. Furthermore, small dots of silver epoxy can be seen on the sample where gold wires were attached for transport measurements (not shown in this work). Right: EDS results for a TaPtTe$_5$ sample. The spectrum confirms that the atomic ratio is close to Ta:Pt:Te = 1:1:5.}\label{characterization_TaPtTe5}
\end{figure}

\section{\texorpdfstring{Nodal-lines in the $k_z=\pm\pi/c$ plane}{}}\label{nodal_line_plane}

\begin{figure}[ht]
        \includegraphics[width=0.48\columnwidth]{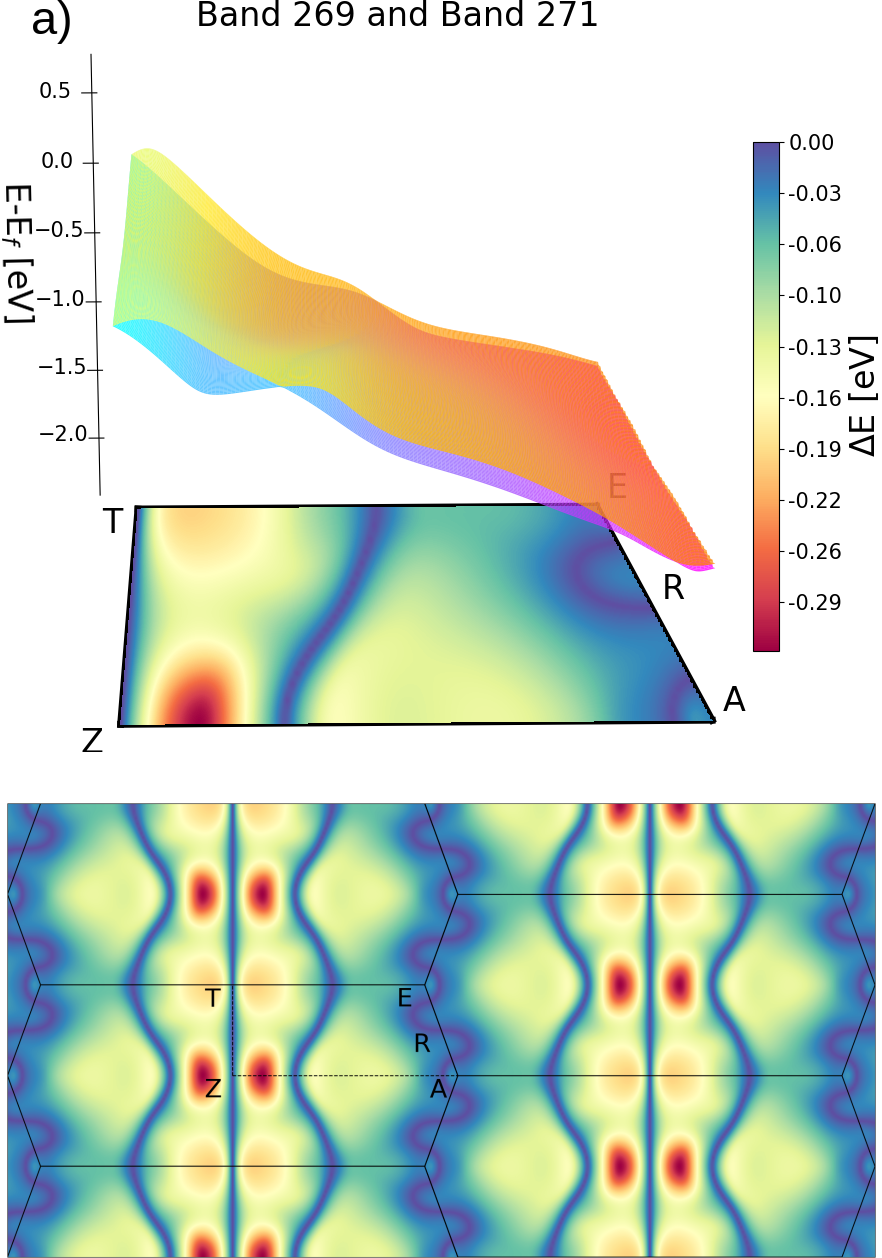}
	\includegraphics[width=0.48\columnwidth]{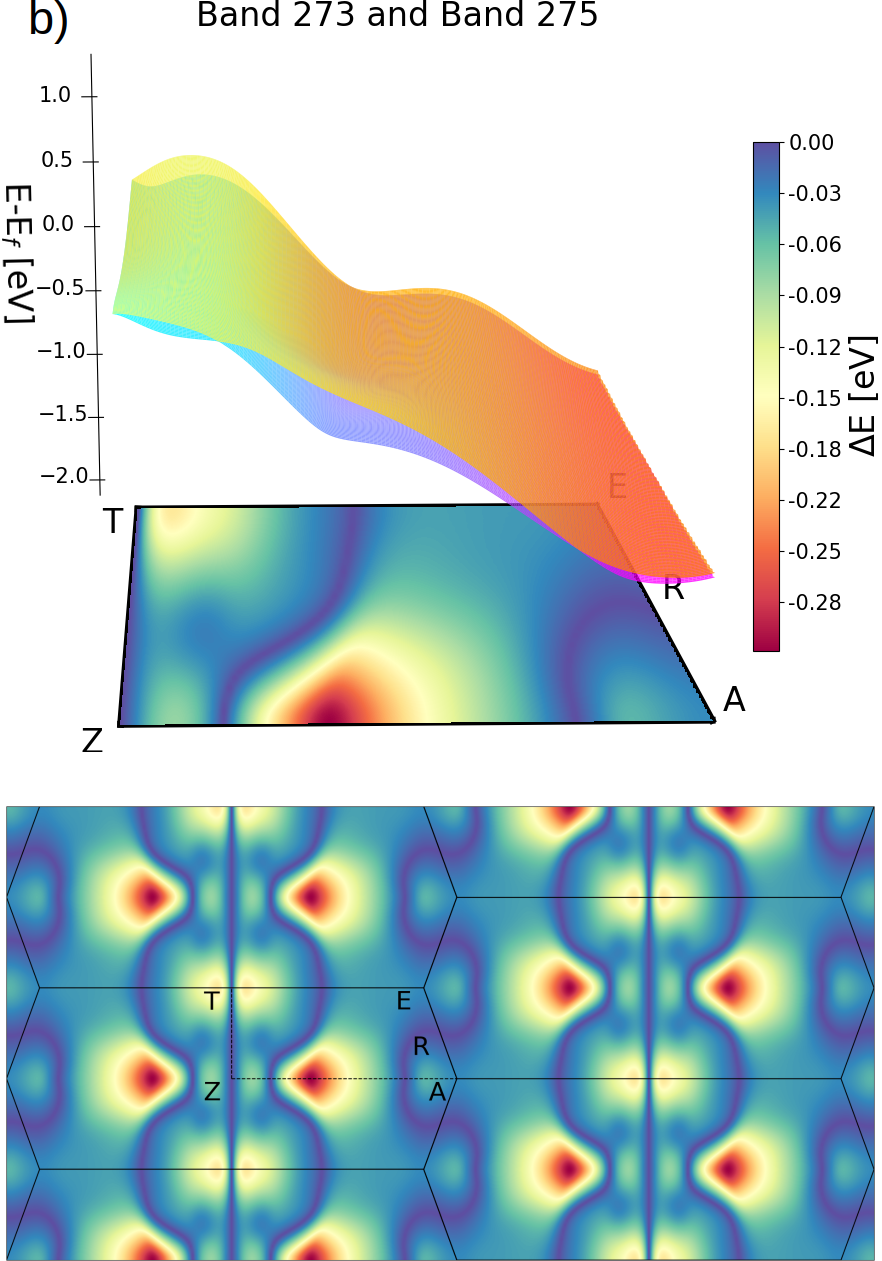}
	\caption{Band structure in TaPtTe$_5$ in the $k_z = \pm\pi/c$ plane including SOC with a) illustrating band 269 and band 271, and b) illustrating band 273 and band 275. The energetic difference between two bands is shown in the contour plots below. For better visualisation, the contour plot is extended to multiple Brillouin zones. One can clearly distinguish three nodal lines in both cases: one straight line in Z-T, one curve that does not intersect any high-symmetry points, and one curve along the Brillouin zone boundary connecting the high-symmetry points in R. Crystal symmetries protect the nodal line along Z-T and the nodal point in R. The numerical results shown here are consistent with those shown by S. Xiao et al. \cite{xiao2022dirac}.}
	\label{TaPtTe5_bandstructure_surface_plot}
\end{figure}

\newpage
\section{DHvA oscillations in the magnetic torque}\label{raw_data_dhva}
\begin{figure}[ht]
  \centering
  \includegraphics[width=\columnwidth]{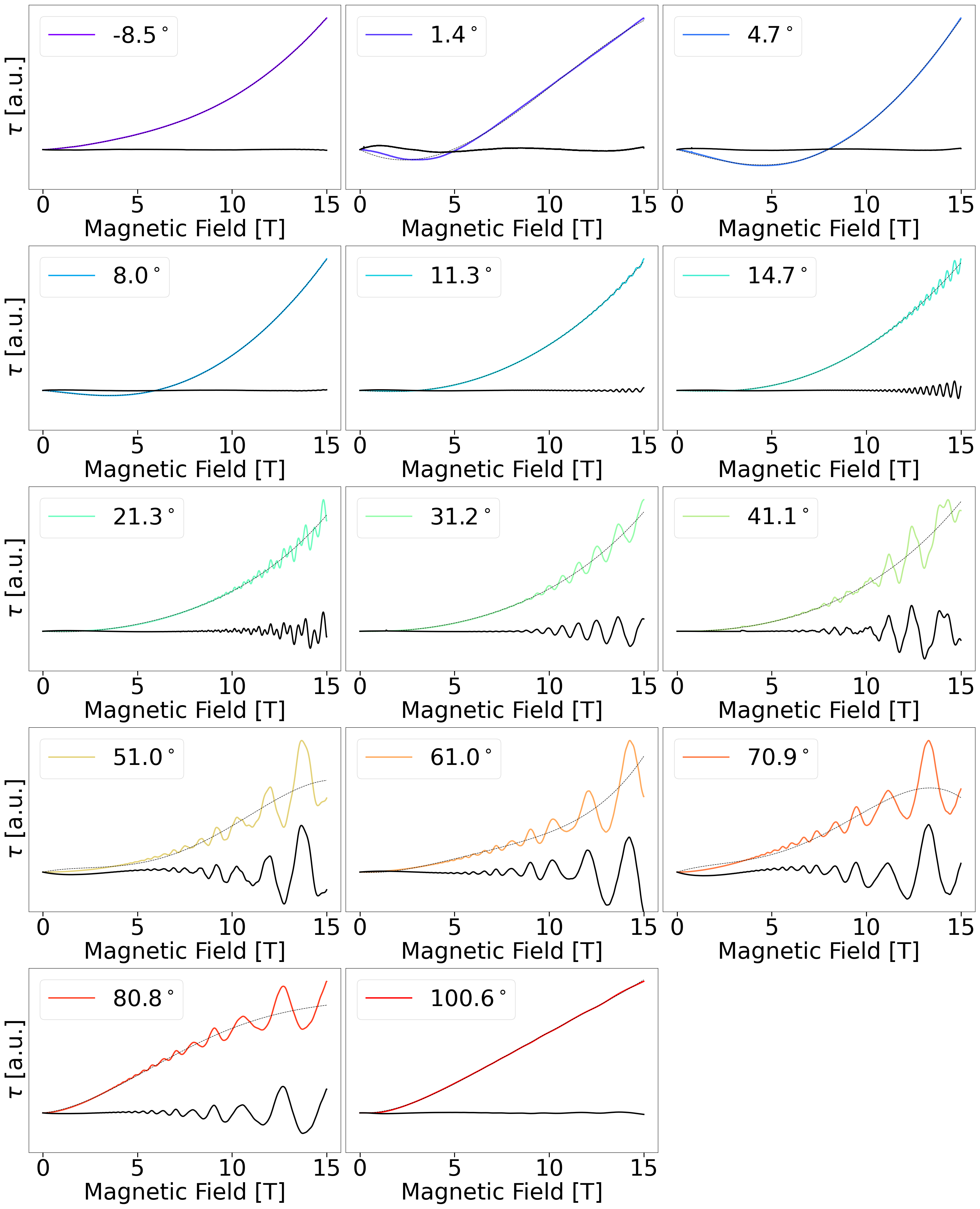}
  \caption{Additional magnetic torque data for a TaPtTe$_5$ sample rotated in the \textit{b}-\textit{c} plane. b). The whole data set was fitted with a fourth-degree polynomial (dashed black line), and the background-subtracted data is given as a solid black line. \SI{0}{\degree} corresponds to the magnetic field being aligned with the \textit{c}-axis. Although performed with the magnetic field in the same plane as given in \autoref{TaPtTe5_overall_torque_individual_b-c_2}, this data set was acquired using a different piezoresistive cantilever which can explain the distinct background signal.}
  \label{TaPtTe5_overall_torque_individual_b-c_1}
\end{figure}

\begin{figure}[ht]
        \includegraphics[width=\columnwidth]{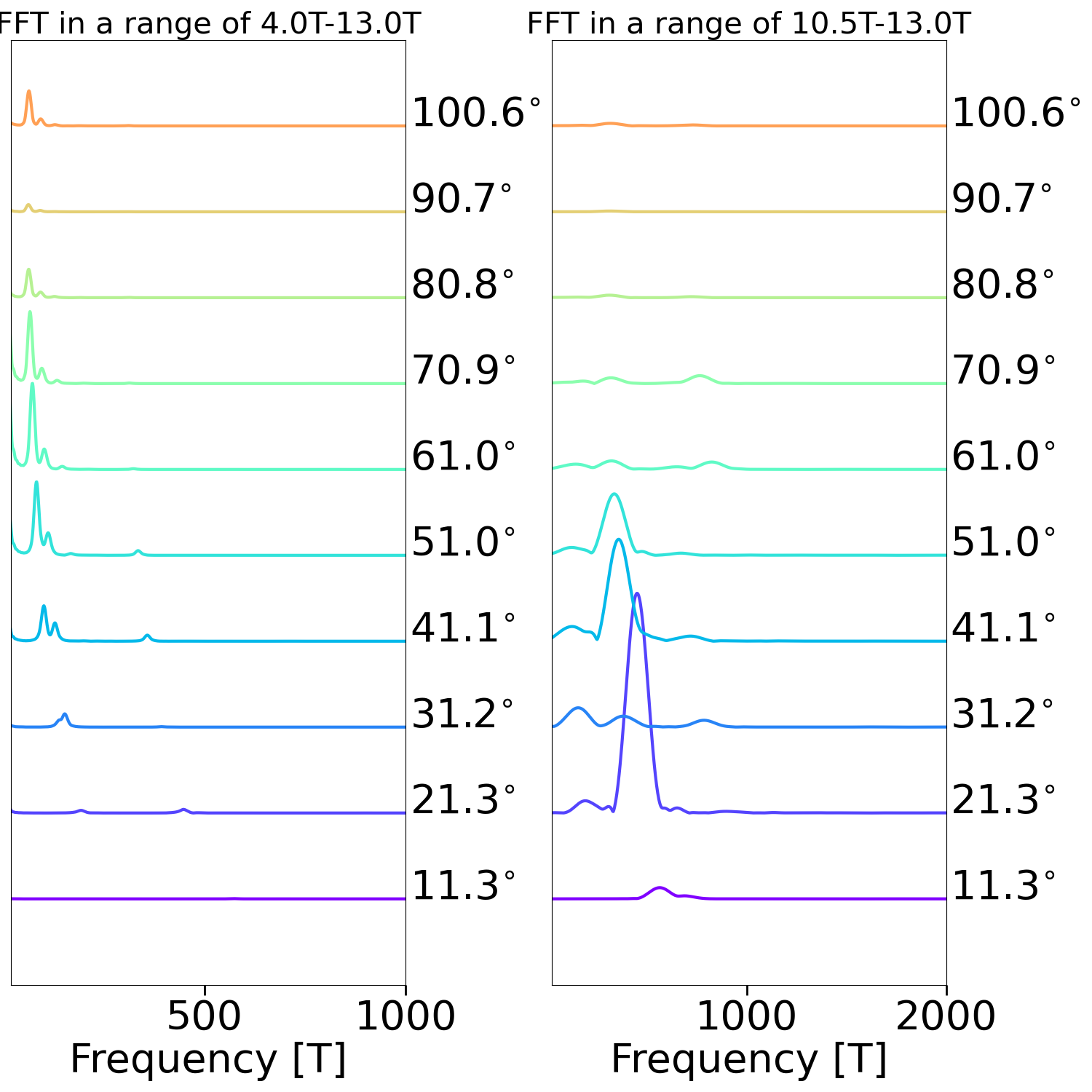}
        \includegraphics[width=\columnwidth]{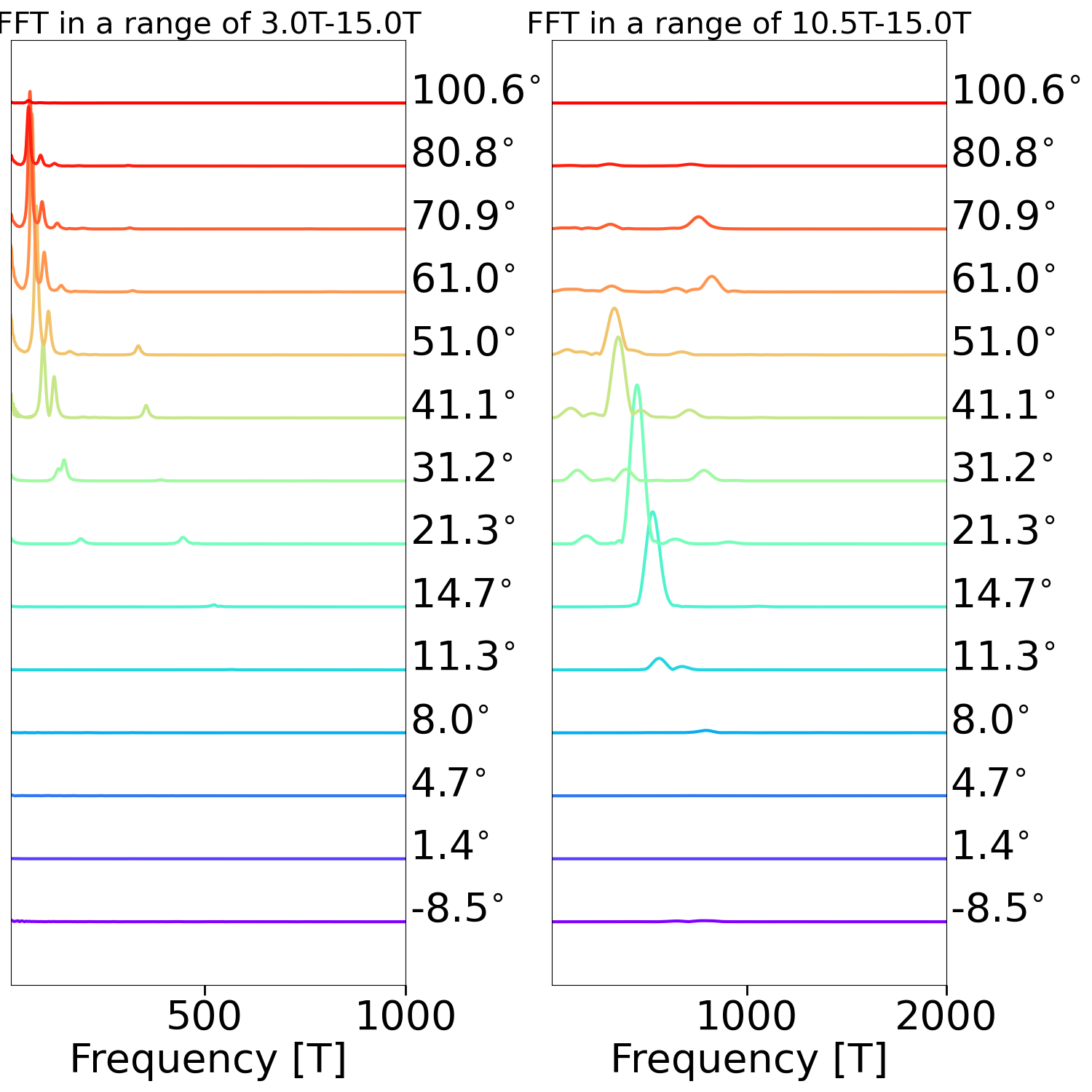}
	\caption{Fast Fourier Transform (FFT) for TaPtTe$_5$ in the \textit{b-c} plane. \SI{0}{\degree} corresponds to the magnetic field being aligned with the \textit{c}-axis. The top row shows data from measurements up to \SI{13}{\tesla} (\autoref{TaPtTe5_overall_torque_individual_b-c_2}), while the bottom row shows data from measurements up to \SI{15}{\tesla} (\autoref{TaPtTe5_overall_torque_individual_b-c_1}).}
     \label{FFT_TaPtTe5_2}
\end{figure}

\begin{figure}[ht]
  \centering
  \includegraphics[width=\columnwidth]{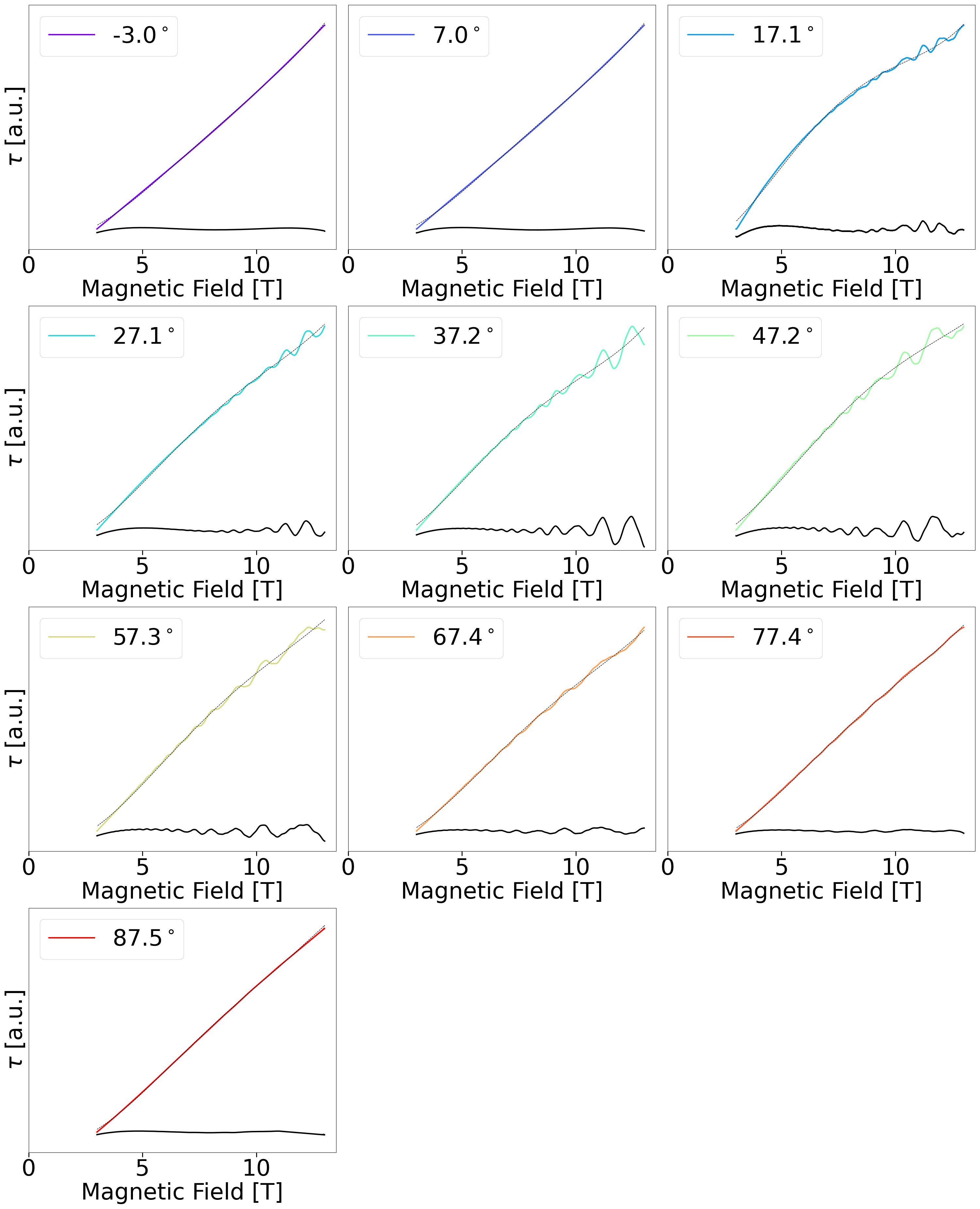}
  \caption{Magnetic torque of a TaPtTe$_5$ sample rotated in the \textit{a}-\textit{b} plane. a). The whole data set was fitted with a fourth-degree polynomial (dashed black line), and the background-subtracted data is given as a solid black line. \SI{0}{\degree} corresponds to the magnetic field being aligned with the \textit{a}-axis.}
  \label{TaPtTe5_overall_torque_individual_a-b}
\end{figure}

\begin{figure}[ht]
  \centering
  \includegraphics[width=\columnwidth]{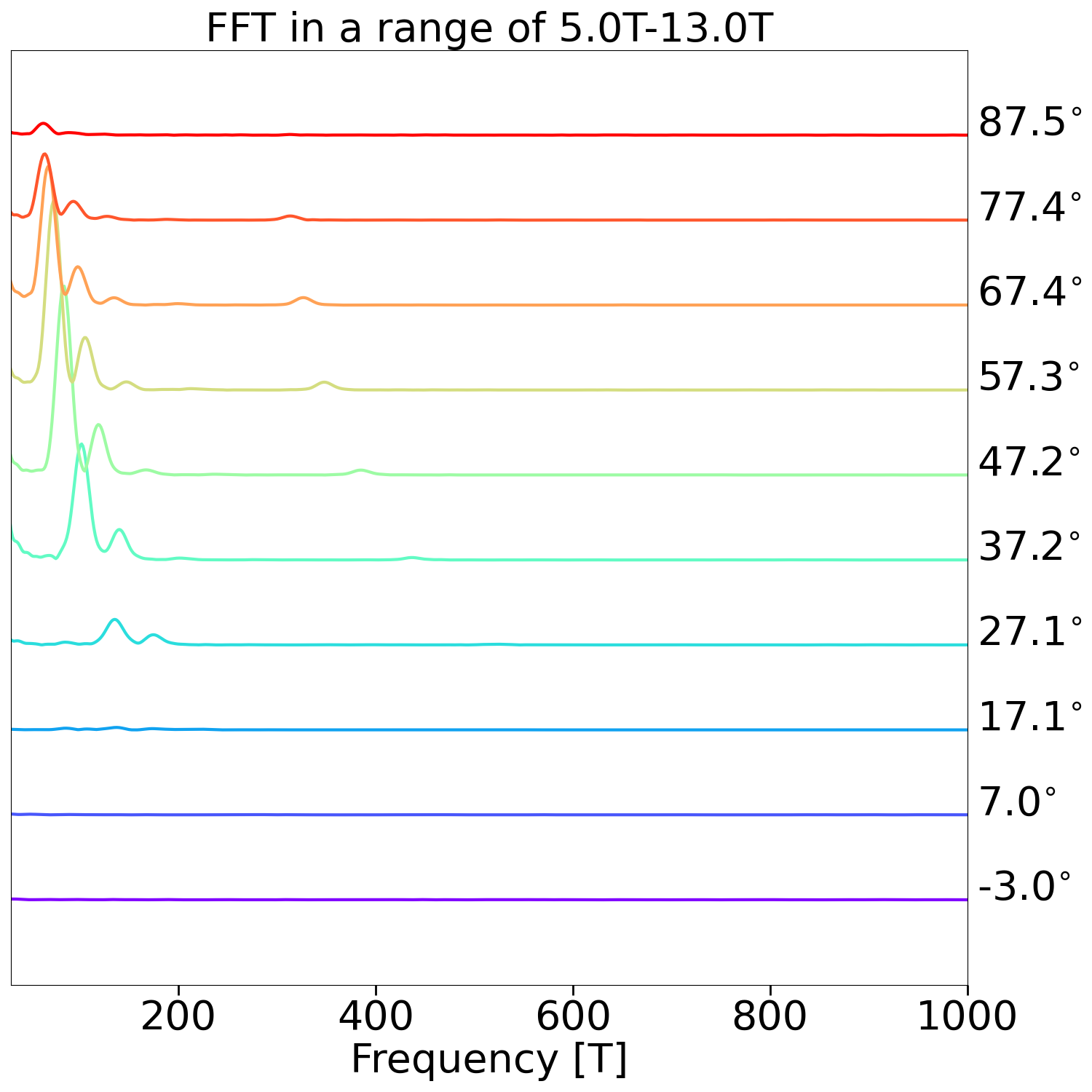}
  \caption{Raw Fast Fourier Transform (FFT) for the data shown in \autoref{TaPtTe5_overall_torque_individual_a-b} for TaPtTe$_5$ in the \textit{a-b} plane. \SI{0}{\degree} corresponds to the magnetic field being aligned with the \textit{a}-axis.}
  \label{FFT_TaPtTe5_1}
\end{figure}

Together with \autoref{TaPtTe5_overall_torque_individual_b-c_2}, the data included in this appendix forms the basis for the frequency peaks shown in \autoref{TaPtTe5_frequencies_torque}. Additional data in the \textit{b}-\textit{c} plane is given in \autoref{TaPtTe5_overall_torque_individual_b-c_1}, and fast Fourier transforms for both, \autoref{TaPtTe5_overall_torque_individual_b-c_2} and \autoref{TaPtTe5_overall_torque_individual_b-c_1}, are included in \autoref{FFT_TaPtTe5_2}. Data for dHvA oscillations in the \textit{a}-\textit{b} plane is given in \autoref{TaPtTe5_overall_torque_individual_a-b} with the respective FFT in \autoref{FFT_TaPtTe5_1}.

\clearpage
\bibliographystyle{unsrt}  
\bibliography{TaPtTe5_QO.bib}

\end{document}